\documentclass[amsmath,amssymb,showpacs,preprint]{revtex4}
\usepackage{graphicx}
\usepackage{epsfig}
\usepackage{graphicx}

\begin{document}

\title{Spatial and temporal variations of the fine-structure constant in Finslerian universe}

\author{Xin Li $^{1,2}$}
\email{lixin1981@cqu.edu.cn}
\author{Hai-Nan Lin $^1$}
\email{linhn@ihep.ac.cn}
\affiliation{$^1$Department of Physics, Chongqing University, Chongqing 401331, China\\
$^2$CAS Key Laboratory of Theoretical Physics, Institute of Theoretical Physics, Chinese Academy of Sciences, Beijing 100190, China}

\begin{abstract}
  Recent observations show that the electromagnetic fine-structure constant, $\alpha_e$, may vary with space and time. In the framework of Finsler spacetime, we propose here an anisotropic cosmological model, in which both the spatial and temporal variations of $\alpha_e$ are allowed. Our model naturally leads to the dipole structure of $\alpha_e$, and predicts that the dipole amplitude increases with time. We fit our model to the most up-to-date measurements of $\alpha_e$ from the quasar absorption lines. It is found that the dipole direction points towards $(l,b)=(330.2^\circ\pm7.3^\circ,-13.0^\circ\pm5.6^\circ)$ in the galactic coordinates, and the anisotropic parameter is $b_0=(0.47\pm 0.09) \times10^{-5}$, which corresponds to a dipole amplitude $(7.2\pm 1.4)\times 10^{-8}$ at redshift $z=0.015$. This is well consistent with the upper limit of the variation of $\alpha_e$ measured in the Milky Way. We also fit our model to the Union2.1 type Ia supernovae, and find that the preferred direction of Union2.1 is well consistent with the dipole direction of $\alpha_e$.
\end{abstract}

\maketitle

\section{Introduction}

The fundamental physical constants play an essential role in the modern physics. It is specially important to test the constancy of the fundamental physical constants both on experiments and theories \cite{Uzan}. The stability of the electromagnetic fine-structure constant $\alpha_e=e^2/\hbar c$ has been extensively tested by various experiments. The Oklo phenomenon is the first experimental approach to test the constancy of $\alpha_e$ \cite{Damour}. The recent analysis on the Oklo phenomenon gives $\Delta\alpha_e/\alpha_e=(3.85\pm5.65)\times10^{-8}$ \cite{Petrov} and $\Delta\alpha_e/\alpha_e=(-0.65\pm1.75)\times10^{-8}$ \cite{Gould}, respectively. The atomic clock data give a strict constraint on the temporal variation of fine-structure constant, i.e., $\dot{\alpha_e}/\alpha_e = (-1.6\pm2.3)\times 10^{-17}\rm yr^{-1}$ \cite{Rosenband}, where the dot denotes the derivative with respect to time. The Planck cosmic microwave background data limit the variation of $\alpha_e$ from $z\approx 1000$ to the present day to be less than approximately $4\times 10^{-3}$ \cite{Planck XVI}. Recently, the observations on quasar absorption spectra show that the fine-structure constant may vary at cosmological scale \cite{Webb,King:2012}. Furthermore, in high-redshift region ($z>1.6$), it has been shown that the variation of $\alpha_e$ is well represented by an angular dipole model pointing to the direction $(l,b)=(330^\circ,-15^\circ)$ in the galactic coordinates with $\sim4.2\sigma$ statistical significance, and the dipole amplitude is about $(0.97^{+0.22}_{-0.20})\times10^{-5}$ \cite{Webb}. Recently, Pinho \& Martins \cite{Pinho} carried out a joint analysis of a larger number of old $\alpha_e$ data \cite{King:2012} and ten more recent measurements of $\alpha_e$ \cite{Molaro,Evans,Whitmore,Ferreira,Songaila,Molaro1,Chand,Agafonova}, and found that the dipolar variation of $\alpha_e$ with amplitude $(0.81\pm0.17)\times10^{-5}$ is still a good fit to the combined data set, while the statistical uncertainty is significantly reduced compared with previous results.

Theoretically, several cosmological models have been proposed to explain the spatial variation of $\alpha_e$, such as models of dark energy (for example, quintessence field or tachyon field) \cite{Copeland}. A more recent analysis of the relation between the dark energy  models and the variation of $\alpha_e$ can be found in Ref. \cite{Martins}. Antoniou \& Perivolaropoulos \cite{Antoniou} found from the Union2 Sne Ia data set that the universe has a preferred direction pointing to $(l,b)=(309^\circ,18^\circ)$.  Later on, Mariano \& Perivolaropoulos \cite{Mariano} showed that the dipole of $\alpha_e$ is approximately aligned with the corresponding dark energy dipole obtained through the Union2 sample. In a recent paper \cite{Finsler dipole}, we proposed an anisotropic cosmological model in Finsler spacetime \cite{Book by Bao}, and showed that both the fine-structure constant and the accelerating cosmic expansion have dipole structure. The fitting to the observational data showed that these two dipole directions are almost aligned. However, the dipole amplitude of $\alpha_e$ in this model is constant, and is in the order of $\sim 10^{-5}$. This conflicts to the upper limit of the variation of $\alpha_e$ measured in the Milky Way, i.e., $|\Delta\alpha_e/\alpha_e|<1.1\times10^{-7}$ \cite{Levshakov}. Webb et al. \cite{Webb:1999} have shown that the variation of $\alpha_e$ mainly comes from high redshift region. We may expect that $\alpha_e$ evolves not only with space, but also with time. In this paper, we will construct a new cosmological model based on the Finsler spacetime. Unlike the previous model, in our new model both the spatial and temporal variations of $\alpha_e$ are allowed.

The rest of the paper is organised as follows. In Section \ref{sec:anisotropy}, we introduce an anisotropic cosmological model in the framework of Finsler spacetime. In Section \ref{sec:field-equation}, we obtain the gravitational field equations and the distance-redshift relation in Finslerian universe. In section \ref{sec:observation}, we use the most up-to-date measurement of $\alpha_e$ and the Sne Ia data to constrain the model parameters. Finally, discussions and conclusions are presented in Section \ref{sec:conclusion}.

\section{Anisotropic universe in Finsler spacetime}\label{sec:anisotropy}

Finsler geometry \cite{Book by Bao} is a generalization of Riemann geometry and includes the latter as its special case. In fact, Finsler geometry is just Riemann geometry without the quadratic restriction. Finsler geometry is based on the so called Finsler structure $F$ defined on the tangent bundle of a manifold $M$, with the property $F(x,\lambda y)=\lambda F(x,y)$ for any $\lambda>0$, where $x\in M$ represents the position and $y$ represents the velocity. The Finslerian metric is given by the second order derivative of $F^2$ with respect to $y$ \cite{Book by Bao},
\begin{equation}
g_{\mu\nu}\equiv\frac{\partial}{\partial
y^\mu}\frac{\partial}{\partial y^\nu}\left(\frac{1}{2}F^2\right).
\end{equation}
Throughout this paper, the indices labeled by Greek letters are lowered and raised by $g_{\mu\nu}$ and its inverse matrix $g^{\mu\nu}$, respectively.
Finslerian metric reduces to Riemannian metric if $F^2$ is quadratic in $y$.

In this paper, we propose a Finsler spacetime that describes the anisotropic evolution of the universe. It is of the form
\begin{equation}\label{FRW like}
F^2=y^ty^t-a^2(t)F_{Ra}^2,
\end{equation}
where
\begin{equation}
F_{Ra}=\sqrt{\delta_{ij}y^iy^j}+b(t)y^z
\end{equation}
is the 3-dimensional Randers space \cite{Randers}, and $a(t)$ is the scale factor of the universe. Hereafter, Greek and Latin indices stand for the 4-dimensional spacetime and 3-dimensional space, respectively. We assume that the Finslerian parameter $b$ is independent of space coordinates, but it is only a function of time. Then, the Killing equations of Randers space $F_{Ra}$ \cite{Finsler PF,Finsler BH} tells us that the translational symmetry of space and the rotational symmetry of $x-y$ plane are preserved. This means that the symmetry of Finsler spacetime (\ref{FRW like}) is the same to that of Bianchi type I spacetime \cite{Bianchi spacetime,Koivisto}. However, we will show in the following that the gravitational field equations of Finsler spacetime (\ref{FRW like}) and Bianchi type I spacetime are different.

The first order variation of Finslerian length gives the geodesic equation which describes the motion of free particles in curved spacetime. The redshift of photon in Finslerian universe can be deduced from the equation of motion of photon. The geodesic equation in Finsler spacetime is given as
\begin{equation}
\label{geodesic}
\frac{d^2x^\mu}{d\tau^2}+2G^\mu=0,
\end{equation}
where
\begin{equation}
\label{geodesic spray}
G^\mu=\frac{1}{4}g^{\mu\nu}\left(\frac{\partial^2 F^2}{\partial x^\lambda \partial y^\nu}y^\lambda-\frac{\partial F^2}{\partial x^\nu}\right).
\end{equation}
It can be easily proven from the geodesic equation (\ref{geodesic}) that the Finsler structure $F$ is a constant along the geodesic. Substituting equation (\ref{FRW like}) into equation (\ref{geodesic spray}), and through a straightforward calculation we obtain
\begin{eqnarray}\label{geodesic spray t}
G^t&=&\frac{1}{2}(a\dot{a}F_{Ra}^2+a^2\dot{b}y^zF_{Ra}),\\
\label{geodesic spray i}
G^i&=&Hy^iy^t+\frac{1}{2}\dot{b}y^t\left(y^i\frac{y^z}{F_{Ra}}+F_{Ra}\bar{g}^{iz}\right),
\end{eqnarray}
where $H\equiv\dot{a}/a$ is the Hubble parameter and $\bar{g}^{ij}$ is the Finslerian metric in Randers space $F_{Ra}$, and the dot represents the derivative with respect to time. In order to get the equation of motion for photon, one should notice that the null condition of photon is given by $F=0$ in Finsler spacetime. Substituting the null condition into the geodesic equation (\ref{geodesic}), we obtain the solution
\begin{equation}
\frac{dt}{d\tau}\propto\frac{1}{a}\exp\left(-b\hat{n}^z\right),
\end{equation}
where $\hat{n}^z$ denotes the unit vector along the $z$-axis.

It yields the formula of redshift $z$
\begin{equation}\label{redshift1}
1+z=\frac{1}{ca}\exp\left(-b\hat{n}^z\right),
\end{equation}
where $c$ is the speed of light at any epoch, and we have assumed that the speed of light at present is $c_0=1$. In Finsler spacetime (\ref{FRW like}), the parameters $a(t)$ and $b(t)$ are constant on local inertial system. We can reparameterize the curve parameter $\tau$ such that $a=1$ on local inertial system. Thus, the speed of light along the radial direction in Finsler spacetime is given by
\begin{equation}\label{SOL}
c_r=\frac{1}{1+b\hat{n}^z}.
\end{equation}
Substituting equation (\ref{SOL}) into equation (\ref{redshift1}), to first order in $b$, we obtain
\begin{equation}\label{redshift2}
1+z=\frac{1}{a}.
\end{equation}

\section{Gravitational field equations and distance-redshift relation}\label{sec:field-equation}

In Finsler geometry, there is a geometrical invariant quantity, i.e., the Ricci scalar. Ricci scalar is related to the second order variation of Finslerian length. In physics, the second order variation of Finslerian length gives the geodesic deviation equation which describes gravitational effect between two particles moving along the geodesics. The analogy between geodesic deviation equations in Finsler spacetime and Riemann spacetime gives the vacuum field equation in Finsler gravity \cite{Finsler BH,Finsler Bullet}, namely, the vanish of Ricci scalar. The Ricci scalar in Finsler geometry is given by \cite{Book by Bao}
\begin{equation}\label{Ricci scalar}
{\rm Ric}\equiv\frac{1}{F^2}\left(2\frac{\partial G^\mu}{\partial x^\mu}-y^\lambda\frac{\partial^2 G^\mu}{\partial x^\lambda\partial y^\mu}+2G^\lambda\frac{\partial^2 G^\mu}{\partial y^\lambda\partial y^\mu}-\frac{\partial G^\mu}{\partial y^\lambda}\frac{\partial G^\lambda}{\partial y^\mu}\right).
\end{equation}
Substituting equations (\ref{geodesic spray t}) and (\ref{geodesic spray i}) into equation (\ref{Ricci scalar}), we obtain
\begin{eqnarray}\label{Ricci scalar1}
F^2{\rm Ric}=-3\frac{\ddot{a}}{a}y^ty^t+(a\ddot{a}+2\dot{a}^2)F^2_{Ra}+(6a\dot{a}\dot{b}+a^2\ddot{b})F_{Ra}y^z-(2\ddot{b}+4H\dot{b})y^ty^tl^z,
\end{eqnarray}
where $l^i\equiv y^i/F_{Ra}$.

In Ref. \cite{Finsler dipole,Finsler BH}, we have proven that the gravitational field equations in Finsler spacetime is of the form
\begin{equation}\label{field equation}
{\rm Ric}^\mu_\nu-\frac{1}{2}\delta^\mu_\nu S=8\pi G T^\mu_\nu,
\end{equation}
where $T^\mu_\nu$ is the energy-momentum tensor. Here the Ricci tensor is defined as \cite{Akbar}
\begin{equation}\label{Ricci tensor}
{\rm Ric}_{\mu\nu}=\frac{\partial^2\left(\frac{1}{2}F^2 {\rm Ric}\right)}{\partial y^\mu\partial y^\nu},
\end{equation}
and the scalar curvature in Finsler spacetime is given as $S=g^{\mu\nu}{\rm Ric}_{\mu\nu}$.
Substituting the equation of Ricci scalar (\ref{Ricci scalar1}) into the field equation (\ref{field equation}), we obtain
\begin{eqnarray}\label{field eq00}
8\pi G T^t_t&=&3H^2+4H\dot{b}l^z+(\ddot{b}+2H\dot{b})\left(\frac{y^t}{aF_{Ra}}\right)^2l^z,\\
\label{field eq0i}
8\pi G T^t_i&=&-\frac{y^t}{F_{Ra}}(\delta^z_i-l^zl_i)(2\ddot{b}+4H\dot{b}),\\
\label{field eqij}
8\pi G T^i_j&=&\left[\frac{2\ddot{a}}{a}+H^2+\left(5H\dot{b}+\frac{3}{2}\ddot{b}\right)l^z\right]\delta^i_j- \left(\frac{y^t}{aF_{Ra}}\right)^2(\ddot{b}+2H\dot{b})(\delta^z_j-2l^zl_j)l^i\nonumber\\
              &&-\left(3H\dot{b}+\frac{1}{2}\ddot{b}\right)(\bar{g}^{iz}l_j+\delta^z_j l^i-l^zl^il_j),
\end{eqnarray}
where $l_i=\bar{g}_{ij}l^j$. The gravitational field equations (\ref{field eq00} -- \ref{field eqij}) imply that the energy-momentum tensor in Finsler spacetime involves shear stress and viscosity that would not appear in Bianchi type I spacetime.

The Planck data give severe constraints on the peculiar velocity \cite{Ade:2014}. Thus, it requires that the energy-momentum tensor in Finsler spacetime (\ref{FRW like}) does not contain viscosity, i.e., $T^t_i=0$. Therefore, it results from equation (\ref{field eq0i}) that
\begin{equation}\label{eq b}
2\ddot{b}+4H\dot{b}=0.
\end{equation}
The solution of equation (\ref{eq b}) is
\begin{equation}\label{eq b1}
\dot{b}=\frac{b_0H_0}{a^2},
\end{equation}
where we have set the integral constant to be $b_0H_0$.

The gravitational field equations (\ref{field eq00} -- \ref{field eqij}) imply that the energy-momentum tensor $T^\mu_\nu$ are a function of both $x$ and $y$. However, $T^\mu_\nu$ in Finslerian manifold has not been well defined. Therefore, noticing that $l^i$ in gravitational field equations are homogeneous function of degree $0$ with respect to the variable $y$, we can set $l^i=\hat{n}^i$ such that the geometrical parts of the gravitational field equations (\ref{field eq00} -- \ref{field eqij}) are only functions of $x$. And, the energy-momentum tensor $T^\mu_\nu$ can be treated as the same in general relativity. Then, by making use of equation (\ref{eq b}), the gravitational field equations (\ref{field eq00}) and (\ref{field eqij}) can be reduced to
\begin{eqnarray}\label{field eq t}
3H^2+4H\dot{b}\hat{n}^z=8\pi G\rho,\\
\label{field eq i}
2\frac{\ddot{a}}{a}+H^2+\frac{4}{3}H\dot{b}\hat{n}^z=-8\pi G p,
\end{eqnarray}
where $\rho$ and $p\equiv(p_x+p_y+p_z)/3$ are the energy density and the mean pressure density of the universe, respectively. We find from equations (\ref{field eq t}) and (\ref{field eq i}) that
\begin{equation}\label{conservation T}
\dot{\rho}+3H(\rho+p)+2\dot{b}\hat{n}^z\left(\rho+\frac{1}{3}p\right)=0.
\end{equation}

In this paper, we assume that the Finslerian universe is made up of pressureless matter component and anisotropic dark energy component. The latter has the equation of state $p_\Lambda=-\rho$. Then, combining equations (\ref{field eq t}) and (\ref{conservation T}), we obtain
\begin{equation}\label{H2}
H^2+\frac{4}{3}H\dot{b}\hat{n}^z=H_0^2\left[\Omega_{m0}a^{-3}\exp(-2b\hat{n}^z)+\Omega_{\Lambda0}\exp(-4b\hat{n}^z/3)\right],
\end{equation}
where $\Omega_{\Lambda0}\equiv8\pi G\rho_{\Lambda0}/(3H_0^2)$ and $\Omega_{m0}\equiv8\pi G\rho_{m0}/(3H_0^2)$ denote the dimensionless density of dark energy and matter at the present epoch, respectively. We assume that our universe is Riemannian at present epoch, and therefore $\Omega_{\Lambda0}=1-\Omega_{m0}$.

The distance-redshift relation in Finslerian universe can be derived from the null condition $F=0$ and equations (\ref{redshift2}) and (\ref{H2}). It is of the form
\begin{equation}\label{lumin dis}
H_0d_L=(1+z)\int_0^z\frac{dz}{(1+b\cos\theta)f(z)},
\end{equation}
where
\begin{equation}
 f(z)\equiv\sqrt{\Omega_{m0}(1+z)^3(1-2b\cos\theta)+(1-\Omega_{m0})(1-\frac{4b}{3}\cos\theta)}-\frac{2}{3}b_0(1+z)^2\cos\theta.
\end{equation}
In the special case $b_0=0$, equation (\ref{lumin dis}) reduces to that of the $\Lambda$CDM model.

\section{Observational constraints}\label{sec:observation}

The variation of the light speed in Finslerian universe leads to the variation of the fine-structure constant. Making use of equations (\ref{SOL}) and (\ref{eq b1}), we obtain the variation of the fine-structure constant to first order in $b$,
\begin{equation}\label{vari alpha}
\frac{\Delta \alpha_e}{\alpha_e}=b\cos\theta,
\end{equation}
where $\theta$ denotes the angle with respect to the $z$-axis, and
\begin{equation}\label{vari_b}
 b(z)=b_0\int_0^z\frac{(1+z)dz}{\sqrt{\Omega_{m0}(1+z)^3+1-\Omega_{m0}}}
\end{equation}
is the dipole amplitude at redshift $z$. Equations (\ref{vari alpha}) and (\ref{vari_b}) mean that $\Delta \alpha_e/\alpha_e$ has a dipole distribution at cosmological scale, and $\Delta \alpha_e=0$ at present epoch. We fit our model to the most completely sample of $\alpha_e$ measured from the quasar absorbtion lines. Our sample contains 303 measurements of $\Delta \alpha_e$ in the redshift range $z\in[0.2223,4.1798]$, among which 293 measurements are taken from Ref. \cite{King:2012}, and the rest 10 measurements are taken from various literatures and are compiled in Ref. \cite{Pinho}. We find the dipole direction pointing towards $(l,b)=(330.2^\circ\pm7.3^\circ,-13.0^\circ\pm 5.6^\circ)$ in the galactic coordinates, with the anisotropic parameter $b_0=(0.47\pm 0.09) \times10^{-5}$. Here, we have fixed $H_0=70.0~\rm km~s^{-1}~Mpc^{-1}$ and $\Omega_{m0}=0.278$, which are derived by fitting to the Union2.1 data set \cite{Suzuki:2012} using the standard $\Lambda$CDM model. Then, we find from the best-fitting result and equation (\ref{vari_b}) that the dipole amplitude at redshift $z=0.015$ is $b({z=0.015})=(7.2\pm1.4)\times10^{-8}$. Our result is well consistent with the upper limit of the variation of $\alpha_e$ measured in the Milky Way, i.e., $|\Delta\alpha_e/\alpha_e|<1.1\times10^{-7}$ \cite{Levshakov}.

To check the reasonability of equation (\ref{vari_b}) in describing the temporal variation of $\alpha_e$, we divide our complete sample into two redshift bins with approximately equal data points in each bin. The low-redshift sample contains 153 data points with redshift $z<1.6$, and the high-redshift sample contains 150 data points with redshift $z>1.6$. We assume that the dipole amplitudes are constants in these two redshift bins, and fit the subsamples to the dipole model with constant amplitude, i.e. $\Delta\alpha_e/\alpha_e=A\cos\theta$. For the low-redshift sample, we find $A=(0.44\pm 0.24) \times10^{-5}$, and the dipole direction points towards $(l,b)=(354.7^\circ\pm 20.0^\circ,-14.2^\circ\pm 15.2^\circ)$. For the high-redshift sample, we find $A=(1.29\pm 0.26) \times10^{-5}$, and the dipole direction points towards $(l,b)=(321.6^\circ\pm 9.0^\circ,-16.5^\circ\pm 6.3^\circ)$. We can see that the dipole directions of both subsamples are consistent with that of the complete sample within $1\sigma$ uncertainty, while the high-redshift subsample has larger dipole amplitude than the low-redshift subsample. Using equation (\ref{vari_b}), we can calculate the average dipole amplitude in this two redshift bins. We obtain $\bar{b}(z_{\rm min}<z<1.6)=(0.48\pm 0.09)\times 10^{-5}$ and $\bar{b}(1.6<z<z_{\rm max})=(1.46\pm 0.28)\times 10^{-5}$, where $z_{\rm min}=0.2223$ and $z_{\rm max}=4.1798$ are the minimum and maximum redshifts of the complete sample, respectively. The average dipole amplitudes in these two redshift bins excellently agree with the results obtained by fitting to the dipole model with constant amplitude. Therefore, equation (\ref{vari_b}) is quantitatively consistent with the observational data.

It has already been noticed that the dipole of Sne Ia is approximately aligned with that of $\alpha_e$, althouth their amplitudes differs by $\sim 2$ orders of magnitude \cite{Mariano}. To compare the variation of $\alpha_e$ with the preferred direction of cosmic accelerating, we also fit our model to the Union2.1  data set \cite{Suzuki:2012}, which is a compilation of 580 Sne Ia in the redshift range $z\in [0.015,1.414]$. The least-$\chi^2$ fit of equation (\ref{lumin dis}) to the Union2.1 data set shows that the preferred direction locates at $(l,b)=(312.8^\circ\pm 19.6^\circ, -11.8^\circ\pm 11.8^\circ)$, and the anisotropic parameter is $b_0=(-2.57\pm 1.15)\times 10^{-2}$. We note that the dipole direction of Sne Ia is consistent with that of $\alpha_e$ within $1\sigma$ uncertainty. However, the anisotropic parameters differ by more than $\sim 3$ orders of magnitude.

\section{Discussion and conclusions}\label{sec:conclusion}

In this paper, we proposed an anisotropic cosmological model in the Finsler spacetime to account for the spatial and temporal variations of the electromagnetic fine-structure constant $\alpha_e$. We obtained the gravitational field equations and the distance-redshift relations in Finslerian universe. In our model, the variation of $\alpha_e$ arises from the variation of light speed. The variation of $\alpha_e$ has the dipole structure, and the dipole amplitude increases with time. We fitted our model to a large set of $\alpha_e$ measurements from the quasar absorbtion lines, and found the dipole direction pointing towards $(l,b)=(330.2^\circ\pm7.3^\circ,-13.0^\circ\pm5.6^\circ)$, with the anisotropic parameter $b_0=(0.47\pm 0.09) \times10^{-5}$. At present epoch, the variation of $\alpha_e$ vanishes, which has been tested with high accuracy by various experiments on earth. Our result is also consistent with the upper limit of $\alpha_e$ variation measured in the Milky Way. What's more, the temporal evolution of $\alpha_e$ is also consistent with the present data. A main shortcoming of our model is that the variation of $\alpha_e$ diverges when $z\rightarrow\infty$, which is of course unreasonable. Our model is only reasonable below a critical redshift. The furthest data point in our sample has redshift $z\sim 4$, and our model can quantitatively match the current data.

We also fit our model to the Union2.1 Sne Ia data, and found that the dipole direction of Union2.1 is approximately aligned with the dipole of $\alpha_x$. However, the anisotropic parameter obtained from Union2.1, $b_0=(-2.57\pm 1.15)\times 10^{-2}$, is more than 3 orders of magnitude larger than that obtained from $\alpha_e$. In fact, Mariano \& Perivolaropoulos \cite{Mariano} have already noticed that the dipole amplitudes of Sne Ia and $\alpha_e$ differ by about 2 orders of magnitude. The existing theoretical models, such as the extended topological quintessence \cite{Mariano} and the Finslerian universe \cite{Finsler dipole}, including the model proposed in this paper, all couldn't reconcile this contradiction. One possible explanation is that the present Sne Ia data are not accurate enough. There are several evidences for this hypotheses. The recent analysis  \cite{JLA HNLin} showed that the dipole directions of two different Sne Ia data sets, Union 2 \cite{Amanullah:2010} and JLA \cite{Betoule}, are inconsistent.  The dipole directions of Union2 derived using two different methods are almost opposite \cite{Chang:2014}. If the dipoles of Sne Ia and $\alpha_e$ have the same origin, e.g., the universe is Finslerian, we may expect that the dipole amplitudes of these two data sets should also be the same.

\begin{acknowledgments}
This work has been supported by the National Natural Science Fund of China (Grant Nos. 11305181 and 11547035), the Fundamental Research Funds for the Central Universities (Grant No. 106112016CDJCR301206), and the Open Project Program of State Key Laboratory of Theoretical Physics, Institute of Theoretical Physics, Chinese Academy of Sciences, China (Grant No. Y5KF181CJ1).
\end{acknowledgments}

\end{document}